\documentclass[10pt,conference]{IEEEtran}

\usepackage{flushend}
\usepackage{times}
\usepackage[cmex10]{amsmath}
\usepackage{amssymb}
\usepackage{epsfig,verbatim}
\usepackage{algorithm}
\usepackage{algpseudocode}

\newcommand{\eqnlabel}[1]{\label{eqn:#1}}
\newcommand{\eqnref}[1]{(\ref{eqn:#1})}

\newcommand\blfootnote[1]{%
  \begingroup
  \renewcommand\thefootnote{}\footnote{#1}%
  \addtocounter{footnote}{-1}%
  \endgroup
}

\providecommand{\Gv}{\mathbf{G}}
\providecommand{\Pv}{\mathbf{P}}
\providecommand{\Sv}{\mathbf{S}}

\providecommand{\yv}{\mathbf{y}}
\providecommand{\ov}{\mathbf{0}}

\providecommand{\Ac}{{\cal A}}
\providecommand{\Cc}{{\cal C}}
\providecommand{\Dc}{{\cal D}}

\providecommand{\Ic}{{\cal I}}

\providecommand{\Yc}{{\cal Y}}

\newtheorem{MyTheorem}{Theorem}
\newcommand{\thmlabel}[1]{\label{thm:#1}}
\newcommand{\thmref}[1]{\ref{thm:#1}}

\newtheorem{definition}{Definition}

\newtheorem{MyLemma}{Lemma}
\newcommand{\lemmalabel}[1]{\label{thm:#1}}
\newcommand{\lemmaref}[1]{\ref{thm:#1}}

\newtheorem{MyProposition}{Proposition}

\title{Capacity-Achieving Rate-Compatible \\ Polar Codes}
\begin{document}



\author{
\IEEEauthorblockN{
              Song-Nam Hong, Dennis Hui and Ivana Mari\'c}
\IEEEauthorblockA{Ericsson Research\\
              200 Holger Way, San Jose, CA, USA\\
              Email: \{songnam.hong, dennis.hui, ivana.maric\}@ericsson.com}
}

\maketitle

\date{}

\blfootnote{This work was presented in part at the $8$th North American School of Information Theory (NASIT 2015), UCSD, La Jolla, CA, Aug. 10-14, 2015.
}
\begin{abstract}
A method of constructing rate-compatible polar codes that are capacity-achieving with low-complexity sequential decoders is presented.
The proposed code construction allows for incremental retransmissions at different rates in order to adapt to channel conditions.
The main idea of the construction exploits certain common characteristics of polar codes that are optimized for a sequence of successively degraded channels. The proposed approach allows for an optimized polar code to be used at every transmission thereby achieving capacity.
Due to the length limitation of conventional polar codes, the proposed construction can only support a restricted set of rates that is characterized by the size of the kernel when conventional polar codes are used. To overcome this limitation,  punctured polar codes which provide more flexibility on block length by controlling a puncturing fraction are considered.  The existence of capacity-achieving punctured polar codes for any given puncturing fraction is shown.  Using punctured polar codes as constituent codes, it is shown that the proposed rate-compatible polar code is capacity-achieving for an arbitrary sequence of rates and for any class of degraded channels.
\end{abstract}


\section{Introduction}
Polar codes, proposed by Arikan \cite{Arikan2009}, achieve the symmetric capacity of the binary-input discrete memoryless channels using a low-complexity successive cancellation (SC) decoder. The finite-length performance of polar codes can be improved by deploying list decoder  enabling polar codes to approach the performance of optimal maximum-likelihood (ML) decoder \cite{TalVardy2011}. Furthermore, a polar code  concatenated with a simple CRC outperforms well-optimized LDPC and Turbo codes even for short block lengths \cite{TalVardy2011}. Due to their good performance and low complexity, polar codes are currently considered for possible use in future wireless communication systems (e.g. 5G cellular systems).

Wireless broadband systems operate in the presence of time-varying channels and therefore require flexible and adaptive transmission techniques. For such systems, hybrid automatic repeat request based on incremental redundancy (HARQ-IR) schemes are often used, where parity bits are sent in an incremental fashion depending on the quality of the time-varying channel.   In  HARQ-IR scheme, a number of parity bits chosen according to a rate requirement,  are sent by the transmitter. IR systems require the use of {\it rate-compatible}  codes typically obtained by puncturing.   For a code to be rate-compatible, the set of parity bits of a higher rate code should be a subset of the set of parity bits of a lower rate code.  This allows the receiver that fails to decode at a particular rate, to request only additional parity bits from the transmitter.  For this reason, there has been extensive research on the construction of rate-compatible Turbo codes and LDPC codes \cite{Hagenauer1988, Rowitch2000, Ha2004,El-Khamy2009,Tsung-Yi2015}.

Although polar codes can achieve the capacity of symmetric  binary-input  channel, their rate-compatible constructions are not in general capacity-achieving. Puncturing of polar codes incurs a rate loss and determining puncturing patterns that result in good performance was considered in \cite{Eslami2011},  \cite{Huawei2014}, \cite{Chen2013}, \cite{WangLiu2014}, \cite{Shin2013}. More recently,  an efficient algorithm for joint optimization of  puncturing patterns and the set of information bits of the code  was proposed and shown to outperform LDPC codes \cite{Miloslavskaya2015}. Because in \cite{Eslami2011}, \cite{Chen2013}, \cite{WangLiu2014}, \cite{Shin2013}, \cite{Miloslavskaya2015} an information set is
optimized according to a puncturing pattern, these
methods cannot be used to design a family of rate-compatible
punctured codes as required for HARQ-IR, where the same
information set (generally optimized for the mother code) should
be used for all punctured codes in the family. In \cite{Huawei2014}, a
heuristic search algorithm was presented to design a good
puncturing pattern for a fixed information set. However, finding an optimal rate-compatible
puncturing pattern with low complexity is still an open problem.

In this paper, we present a family of rate-compatible polar codes that are capacity-achieving. The main idea of our construction exploits certain common characteristics of polar codes that are optimized for a sequence of successively degraded channels. In our approach, we construct a concatenated polar code  that is decoded by a sequence of parallel polar decoders. We refer to this construction as {\it parallel concatenated polar codes}.  The proposed code construction  allows for incremental retransmissions at different transmission rates and can therefore  be used for HARQ-IR.  Furthermore, an optimized polar code is used at every retransmission thereby achieving capacity, for any class of degraded channels. In a work developed independently and in parallel to ours, a similar idea was introduced by Li et. al. in \cite{LiTse2015}. Due to the length limitation of polar codes, our proposed construction of rate-compatible polar codes can achieve the capacity only for a sequence of rates that satisfy a certain relationship (as specified in Theorem~\thmref{theoremnopuncturing}). In order   to support any arbitrary sequence of rates, we present capacity-achieving  punctured polar codes which can provide more flexibility on block length by controlling a puncturing fraction. Using such punctured polar codes, we  show that the proposed rate-compatible polar code is capacity-achieving for an arbitrary sequence of rates.

The paper is organized as follows.  Section \ref{Preliminaries} gives the definition of rate-compatible codes  and  preliminaries on polar codes. In Section \ref{PuncturedPolarCode}, we describe the capacity-achieving punctured code. Section \ref{MainIdea}  presents the main capacity result and the idea of our code design. The precise code construction and decoding are presented in Section \ref{CodeConstruction}.  Comparison of the proposed approach to random puncturing is given in Section \ref{SimulationResults}. Proofs are given in Section \ref{Proofs}. Section \ref{Conclusion} concludes the paper.

\section{Preliminaries} \label{Preliminaries}

In this section, we provide some basic definitions and background  that will be used in the sequel.

\subsection{Rate-Compatible Codes} \label{RateCompatibleCodes}
We start with a  definition of  rate-compatible codes.  To simplify notation, let $[K] \triangleq \{1,2,\cdots,K\}$ for any positive integer $K$ and let $a^n=(a_1, \ldots, a_n)$ denote a vector of length $n$. Given a fixed number of information bits $k$, a family of codes, or code family, $\Cc=\{\Cc_{1}^{\bar{n}_1}, \Cc_{2}^{\bar{n}_2}, \cdots, \Cc_{K}^{\bar{n}_K}\}$ with respective block lengths $\bar{n}_1 < \bar{n}_2 < \cdots <\bar{n}_K$ and corresponding rates $R_1 > R_2 > \cdots > R_K$, where $R_i = k/\bar{n}_i$, is said to be {\it rate-compatible} if there exist a sequence of encoding functions $\{\bar{e}_i(\cdot)\}_{i \in [K] }$, where $\bar{e}_i:\{0,1\}^k \rightarrow \{0,1\}^{\bar{n}_i}$ is the corresponding encoding function of $\Cc_i$ for $i \in [K]$, and a sequence of projection operators $\{\pi_i(\cdot)\}_{i \in [K-1]}$, where $\pi_i:\{0,1\}^{\bar{n}_{i+1}}\rightarrow\{0,1\}^{\bar{n}_{i}}$ simply takes $\bar{n}_{i}$ of the $\bar{n}_{i+1}$ coordinates of its input as output, such that for each $i \in [K-1]$,
\begin{equation} \eqnlabel{RCcondition}
\bar{e}_i(u^k) = \pi_{i}(\bar{e}_{i+1}(u^k)),
\end{equation}
for every possible information block $u^k\in\{0,1\}^k$. We refer to such a sequence of encoding functions as a sequence of {\it nested encoding functions}.
Obviously, any subfamily of $\Cc$ denoted by $\Cc'=\{\Cc_{i_1}^{\bar{n}_{i_1}}, \Cc_{i_2}^{\bar{n}_{i_2}}, \cdots, \Cc_{i_j}^{\bar{n}_{i_j}}\}$, for any $1 \le i_1<i_2, \ldots < i_j \leq K$,  is rate compatible if $\Cc$ is rate compatible.

Condition \eqnref{RCcondition} assures that the set of parity bits of a higher rate code is a subset of the set of parity bits of a lower rate code and therefore the code can be used for HARQ-IR. Specifically, during the $i$-th transmission, the transmitter can send $e_i(u^k) \triangleq \pi^\perp_{i-1}(\bar{e}_i(u^k))$ over the channel, for any given information block $u^k \in \{0,1\}^k$, where $\pi^\perp_{i}:\{0,1\}^{\bar{n}_{i+1}}\rightarrow\{0,1\}^{\bar{n}_{i+1}-\bar{n}_{i}}$ denotes the projection operator orthogonal to $\pi_{i}(\cdot)$ that takes the other $\bar{n}_{i+1}-\bar{n}_{i}$ coordinates of $u^k$ not taken by $\pi_{i}(u^k)$ as output, with $\bar{n}_0 \triangleq 0$ and $\pi^\perp_{0}(u^k) \triangleq u^k$ being the identity mapping.  We let $e_i \triangleq \pi^\perp_{i-1} \circ \bar{e}_i$ and refer to $\{e_i \}_{i\in[K]}$ as a sequence of {\em incremental encoding functions}.  By definition, each family of rate-compatible codes has at least one associated sequence of incremental encoding functions.

It follows immediately that a family of {\it linear} codes ${\bf\cal C}=\{\Cc_{1}^{\bar{n}_1}, \Cc_{2}^{\bar{n}_2}, \cdots, \Cc_{K}^{\bar{n}_K}\}$ is rate-compatible if and only if there exists a sequence of generator matrices $\{{\bf G}_i \}_{i \in [K]}$, each corresponding to a member in ${\bf\cal C}$, such that the columns of ${\bf G}_i$ is a subset of those of ${\bf G}_{i+1}$ for every $i \in [K-1]$.  Accordingly, we refer to $\{{\bf G}_i \}_{i \in [K]}$ as a sequence of {\it nested generating matrices}.

\begin{MyLemma}\lemmalabel{RateLengthRelation}
	The sets of rates $\{R_i\}_{i\in[K]}$ and incremental block lengths $\{n_i\}_{i\in[K]}$ of any rate-compatible code family of size $K$ must satisfy the following equivalent conditions:
	\begin{align}
&\mbox{(a):} \hspace{0.3cm} R_i = \frac{R_1}{1 + \sum_{j=2}^i \frac{n_j}{n_1}},  \hspace{0.3cm} \forall i \in \{2,3,\cdots,K\}, \mbox{ and} \eqnlabel{RateRelatedToLength} \\
&\mbox{(b):} \hspace{0.3cm} n_i = R_1 \left(\frac{1}{R_i} - \frac{1}{R_{i-1}}\right) n_1, \hspace{0.3cm} \forall i \in \{2,3,\cdots,K\}.
\eqnlabel{LengthRelatedToRate}
	\end{align}
\end{MyLemma}
\begin{IEEEproof}
The proof is is given in Section~\ref{proof:lemma1}.
\end{IEEEproof}

Let $W_1 \succeq W_2 \succeq \ldots \succeq W_K $ denote a sequence of successively degraded discrete memoryless channels (DMC) with a common input alphabet ${\cal X}$, respective output alphabets $\{{\cal Y}_i\}_{i\in[K]}$, respective transition probability distributions $\{W_i(y|x) \}_{i \in [K]}$ where $x \in {\cal X}$ and $y \in {\cal Y}_i$, and respective capacities $I(W_1) > I(W_2) > \ldots > I(W_K)$.  A sequence of rate-compatible code families, ${\bf\cal C}_m =\{\Cc_{1}^{\bar{n}_{1,m}}, \Cc_{2}^{\bar{n}_{2,m}}, \cdots, \Cc_{K}^{\bar{n}_{K,m}}\}$ for $m\in \mathbb{N}$, designed for a monotonically increasing sequence of information block sizes $\{k_m\}_{m\in \mathbb{N}}$, is said to be {\it capacity-achieving with respect to $\{W_i\}_{i \in [K]}$ } if, for every $m$, there exist a sequence of decoding functions $\{d_{i,m}(\cdot)\}_{i \in [K] }$, where $d_{i,m}:{\cal Y}^{\bar{n}_{i,m}} \rightarrow \{0,1\}^{k_m}$, and a corresponding sequence of nested encoding functions $\{e_{i,m}(\cdot)\}_{i \in [K] }$ such that for any $\epsilon>0$, we have, for every $i \in [K]$,
\begin{align}
&R_{i,m} \triangleq k_m/\bar{n}_{i,m} > I(W_i) - \epsilon, \text{ and} \eqnlabel{RateCloseToCapacity}\\
&\text{Pr}(u^{k_m} \neq d_{i,m}(y^{\bar{n}_{i,m}})) < \epsilon, \eqnlabel{ProbCloseToZero}
\end{align}
for the joint probability distribution given by $p(u^{k_m},y^{\bar{n}_{i,m}}) = 2^{-{k_m}} W_i^{\bar{n}_{i,m}}(y^{\bar{n}_{i,m}}|e_i(u^{k_m}))$, where $W_i^{n}(y^{n}|x^n)\triangleq \prod_{l=1}^{n}W_i(y_l|x_l)$,
for all sufficiently large $m$.

In this paper, we focus on binary-input discrete memoryless channels (B-DMC) with input alphabet ${\cal X}=\{0,1\}$ and any output alphabet $\Yc$.   We denote the transition probabilities of any B-DMC $W$ by $W(y|0)$ and $W(y|1)$ for all $y \in \Yc$.

%
%
%
%

\subsection{Polar Codes} \label{Polarcodes}

Let $2^\mathbb{N}$ denote the set of powers of two.  For any $n \in 2^\mathbb{N}$, let $\Pv_n \triangleq \Pv_2^{\otimes \log(n)}$ denote the rate-one generator matrix of all polar codes with block length $n$, where $\Pv_2$ is the 2-by-2 Arikan kernel \cite{Arikan2009}.  As shown in \cite{Arikan2009}, as $n$ increases, under successive cancellation (SC) decoding, a fraction of the rows of $\Pv_n$ leads to good bit-channels suitable for carrying information bits while the rest forms bad bit-channels whose input should be frozen to known values, assumed to be zero in this paper.  Thus a polar code of length $n$ is completely determined by the rate-one generator matrix $\Pv_{n}$ and the information set $\Ac$ that specifies the set of good bit-channel locations, with the rate of the code given by the ratio $R = |\Ac|/n$.  Let $\Cc(n, R, \Ac)$ denote a polar code of rate $R$ with information set $\Ac$.  Also let $\Pv_{n}^{\Ac}$ denote its $|\Ac| \times n$ generator matrix formed by only taking the rows of $\Pv_{n}$ that correspond to $\Ac$.  The specific orderings of rows within $\Pv_{n}^{\Ac}$ is unimportant in the following discussion.

Given a finite sequence of channels, $\{W_i\}_{i\in[K]}$, a sequence of polar codes $\{\Cc(n, R_i, \Ac_i)\}_{i\in[K]}$ of a common block length $n$ with different rates $R_1 > R_2 > \cdots > R_K$, each corresponding to one channel, can be obtained by selecting a different information set $\Ac_i$ for each channel $W_i$. We refer to a sequence of polar codes $\{\Cc(n, R_i, \Ac_i)\}_{i\in[K]}$ having the same block length $n$ as a sequence of {\it nested polar codes} if their respective information sets are nested, i.e.
 \begin{equation} \eqnlabel{degradedset}
 \Ac_1 \supseteq \Ac_2 \supseteq \ldots \supseteq \Ac_K.
 \end{equation}
For given  nested information sets $\Ac_{1} \supseteq \Ac_{2} \supseteq \cdots \supseteq \Ac_{K}$, we define an $|\Ac_1| \times n$ generator matrix $\Pv_{n}^{(\Ac_{1},\Ac_{2},\cdots,\Ac_{K})}$ with a partial ordering of rows according to $\{\Ac_{i}\}$ as
\begin{equation} \eqnlabel{StackedPolarGenMatrix}
\Pv_{n}^{(\Ac_{1},\Ac_{2},\cdots,\Ac_{K})} = \left[\begin{array}{c} \Pv_{n}^{\Ac_{K}} \\ \Pv_{n}^{\Ac_{K-1}\setminus \Ac_K} \\ \vdots \\  \Pv_{n}^{\Ac_{1}\setminus \Ac_2} \end{array} \right].
\end{equation}
Note that the only difference between $\Pv_{n}^{\Ac_{1}}$ and $\Pv_{n}^{(\Ac_{1},\Ac_{2},\cdots,\Ac_{K})}$ is that the ordering of rows is more specifically defined in the latter.
For any index set $\Dc \subseteq [n]$, let $\Pv_{n}^{(\Ac_{1},\Ac_{2},\cdots,\Ac_{K})}(\Dc)$ denote a submatrix of $\Pv_{n}^{(\Ac_{1},\Ac_{2},\cdots,\Ac_{K})}$ consisting of rows whose indices belong to $\Dc$.

Given a channel $W_i$, $i \in [K]$, we define the information set for a fixed $\epsilon$ as
	\begin{equation} \eqnlabel{A}
 	\Ac_{i,n}(\epsilon)= \{ j \in [n]: Z(W^{(j)}_{i,n}) \le \epsilon \},
 	\end{equation}
 	 where $W^{(j)}_{i,n}$ denotes the $j$-th bit-channel resulted from a polar code with block length $n$ applied to $W_i$,  and $Z(W)$ denotes the Bhattacharyya parameter of bit-channel $W$ given by
 	 	$Z(W)\triangleq \sum_{y \in \Yc} \sqrt{W(y|0)W(y|1)}$.
 The following result follows directly from \cite[Lemma 4.7]{KoradaPhD2009}.
 \begin{MyLemma}\lemmalabel{Korada} \cite[Lemma 4.7]{KoradaPhD2009}
 	Given a sequence of successively degraded channels $W_1 \succeq W_2 \succeq \ldots \succeq W_K$,  $\{\Cc(n, R_{i,n} \Ac_{i,n}(\epsilon))\}_{i\in[K]}$ is a sequence of nested polar codes for any $\epsilon$.
 \end{MyLemma}

	The  set of good bit-channels  can alternatively be defined   in terms of the symmetric capacity $I(W)$.
 When clear from the context, the dependency of $R_i$, $\Ac_i$ on $\epsilon$  will be omitted.

\begin{definition}
We refer to a family of rate compatible codes $\Cc=\{\Cc_{1}^{\bar{n}_1}, \Cc_{2}^{\bar{n}_2}, \cdots, \Cc_{K}^{\bar{n}_K}\}$ as a family of {\em rate-compatible polar codes}  when a (punctured) polar code is used in every transmission, where a puncturing can be applied to a polar code for the purpose of length adaption. More precisely, $\Cc$ is a family of rate-compatible polar codes if there exists a sequence of incremental encoding functions $\{e_i(\cdot)\}_{i \in [K] }$ where each $e_i(\cdot)$ can be implemented by the encoder of a polar code and possibly with a puncturing.
\end{definition}

 Relation \eqnref{degradedset} is a key property that we will use for constructing rate-compatible polar codes.  However, since nested polar codes have the same block length and varying information block sizes, they cannot directly be used  for HARQ-IR, which in contrast assumes a fixed information block size and allows varying block lengths to achieve different rates with a given error probability tolerance.  To obtain a family of (capacity-achieving) rate-compatible codes, we need to construct multiple sequences of nested polar codes as described later.

\section{Punctured Polar Codes} \label{PuncturedPolarCode}

Polar codes with arbitrary lengths can be obtained by puncturing. A polar code of length $n$ is punctured by removing a set of $s$ columns from its generator matrix, which has the effect of reducing the codeword length from $n$ to $n-s$. We let $\alpha = s/n$ denote a {\em puncturing fraction}.
It is assumed that the receiver knows the locations of the punctured bits, and the decoder estimates both the punctured and transmitted symbols during decoding. The punctured polar code can be encoded and decoded in a similar way as the conventional polar codes.
Formally, a punctured polar code of post-puncturing block length $n$ is characterized by its "mother" (unpunctured) polar code of block length $n_u \in 2^\mathbb{N}$ and a puncturing pattern $p^{n_u} = (p_1,p_2,\cdots,p_{n_u}) \in \{0,1\}^{n_u}$ with $p_i=0$ indicating that the $i$th coded bit is punctured and thus not transmitted.
For a given $p^{n_u}$, let $\pi_{p^{n_u}}:{\cal Y}^{n_u} \rightarrow {\cal Y}^{n}$ be a projection operator that copies $n=w_{H}(p^{n_u})$ coordinates of its input as its output based on the puncturing pattern specified by $p^{n_u}$, where $w_{H}(p^{n_u})$ denotes the number of ones in $p^{n_u}$, i.e. $y^n = \pi_{p^{n_u}}(y^{n_u})$ containing the coordinates of $y^{n_u}$ corresponding to the locations of ones in $p^{n_u}$.
The notion of bit-channels in conventional polar codes can be extended to punctured polar codes in a straightforward manner as follows.  For a given (unpunctured) polar code of block length $n_u \in 2^\mathbb{N}$ and puncturing pattern $p^{n_u}$, we define the transition probability of the $i$-th bit channel of the corresponding punctured polar code as
\begin{eqnarray}
\lefteqn{W^{(i)}(y^n,u^{i-1},p^{n_u}|u_i)} \nonumber \\
& & = \frac{1}{2^{n_u-1}} \sum_{u_{i+1}^{n_u}} \sum_{y^{n_u} \in \pi^{-1}_{p^{n_u}}\left(\{y^n\}\right)} W^{n_u}(y^{n_u}| u^{n_u}{\Pv}_{n_u}),   \eqnlabel{BitChannelPuncturedPolarCode}
\end{eqnarray}
where
\begin{equation}
W^{n_u}(y^{n_u}|x^{n_u}) = \prod_{j \in [n_u]} W(y_j|x_j),
\end{equation} and where
$\pi^{-1}_{p^{n_u}}(S) \triangleq \{y^{n_u} \in {\cal Y}^{n_u}:\pi_{p^{n_u}}(y^{n_u}) \in S\}$ denotes the inverse image of $\pi_{p^{n_u}}(\cdot)$.
Based on \eqnref{BitChannelPuncturedPolarCode}, the information set of a punctured polar code can be defined in a similar manner as in \eqnref{A}.

Let $\Cc(n, R, \Ac, p^{n_u})$ denote a punctured polar code of (post-puncturing) block length $n$, rate $R$, information set $\Ac$, and a puncturing pattern $p^{n_u}$.  Similarly, one can let $\Pv_{n,p^{n_u}}$ denote the matrix obtained by removing the columns of $\Pv_{n_u}$ according to the locations of zeros in the puncturing pattern $p^{n_u}$, and for any nested information sets $\Ac_{1} \supseteq \Ac_{2} \supseteq \cdots \supseteq \Ac_{K}$, one can define $\Pv_{n,p^{n_u}}^{(\Ac_{1},\Ac_{2},\cdots,\Ac_{K})}$ in the same manner as in \eqnref{StackedPolarGenMatrix}.
However, in most cases of our discussion below, there is only one puncturing pattern for each block length $n$, and therefore we will omit  $p^{n_u}$ and use same notations and similar terminologies for both punctured and unpunctured polar codes whenever it is clear from context.


In the proposed scheme that will be explained in Section~\ref{MainIdea}, any good punctured (or shortened) polar code can be used for length flexibility and it is left for a future work to design a good punctured polar code for short block lengths. In order to show that the proposed rate-compatible polar codes achieve the capacity, we will use a capacity-achieving punctured polar code for any desired puncturing fraction, whose existence is shown in the following theorem.
\begin{MyTheorem}\label{thm:punc}
	Consider any B-DMC $W$ with $I(W) > 0$. For any fixed $R < I(W)$, $\beta <\frac{1}{2} $,   and puncturing fraction $\alpha \in (0,1)$, there exists a sequence of punctured polar codes, each with respective block length $n = \lfloor(1-\alpha)2^m\rfloor$ and associated information sets $\Ac_{m} \subset [2^m]$, $m  \in \mathbb{N}$, such that $|\Ac_{m}| \geq \lfloor2^m (1-\alpha)\rfloor R = n R$ and
	\begin{equation}
	P_{e,j,m} \leq O(2^{-2^{m\beta}}) = O(2^{-n^{\beta}}), \eqnlabel{BitErrorCondition}
	\end{equation} for all $j\in \Ac_{m}$ and for all $m \in \mathbb{N}$, where $P_{e,j,m}$  denote the error probability of $j$-th polarized bit channel of the $m$-th punctured polar code.
\end{MyTheorem}

\begin{IEEEproof}
The proof is given in Section~\ref{subsec:thm1proof}.
\end{IEEEproof}

\section{Main Results} \label{MainIdea}

We state two theorems that are the main results of this paper.
\begin{MyTheorem} \thmlabel{capacitytheorem}
For any sequence of successively degraded channels $W_1 \succeq W_2 \succeq \ldots \succeq W_K $, there exists a sequence of rate-compatible polar code families that is capacity-achieving.
\end{MyTheorem}
\begin{IEEEproof}
The proof is given in Section~\ref{subsec:thm2proof}.
\end{IEEEproof}

In case puncturing is not used, the constraint on a polar code block length $n=2^l$, $l \in \mathbb{N}$ reduces the set of rates that can be supported by the proposed coding scheme. In particular, we have the following:
\begin{MyTheorem}\thmlabel{theoremnopuncturing}
For any sequence of successively degraded channels $W_1 \succeq W_2 \succeq \ldots \succeq W_K $ with corresponding symmetric capacities $I(W_1) > I(W_2) > \ldots > I(W_K) > 0$, there exists a sequence of rate-compatible (non-punctured) polar code families that is capacity-achieving  if and only if, for each $i \in \{2, \ldots, K\}$,
\begin{equation}\eqnlabel{Iassumption}
I(W_i)=\frac{I(W_1)}{1 + \sum_{j=2}^i 2^{\ell_j}},
\end{equation}
for some $\ell_j \in \mathbb{Z}$.
\end{MyTheorem}
\begin{IEEEproof}
The proof is given in Section~\ref{subsec:thm3proof}.
\end{IEEEproof}

In the next section, we explain our main idea for the construction of a family of rate-compatible polar codes and also provide a simple example for $K=3$. A general code construction will be explained in Section~\ref{CodeConstruction}.
\subsection{Main Idea}

Before stating the precise construction, we first describe our  idea on how to design the  family of rate-compatible polar codes that are capacity-achieving.

\begin{figure}[t]
  \centering
   \includegraphics[height=1.2in, width=3.2in]{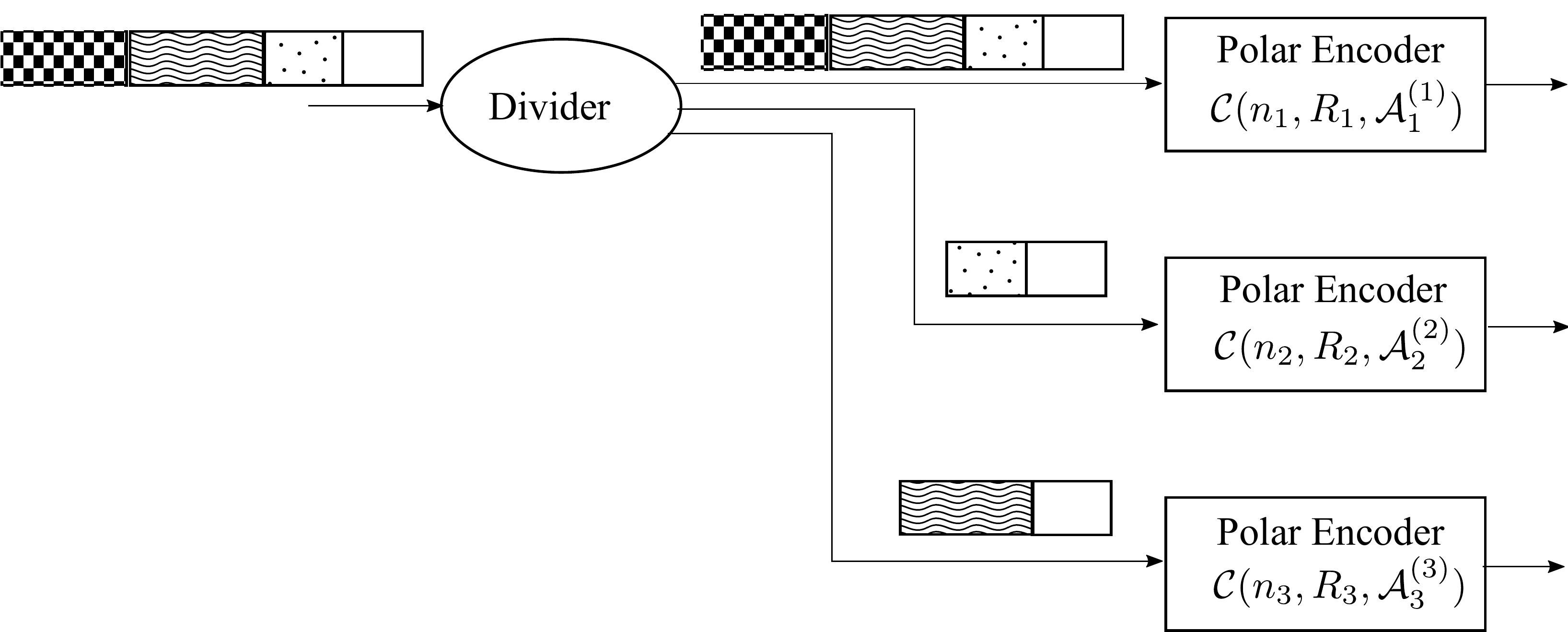}
  \caption{Encoder structure of rate-compatible polar codes for $K=3$.}
       \label{Encoder}
  \end{figure}

To transmit $k$ information bits over $K$ channels $W_1 \succeq W_2 \succeq \ldots \succeq W_K $  at rates $R_1 > \ldots > R_K$, we generate $K$  (punctured) polar codes $\Cc (n_i, R_i, \Ac_i^{(i)})$, where $n_i$ is the block length of the $i$-th transmission and is chosen such that
\begin{equation}
R_i= \frac{k}{\sum_{j=1}^i n_j} = \frac{k}{\bar{n}_i},  \eqnlabel{EffectiveRates}
\end{equation}
where $\bar{n}_i \triangleq \sum_{j=1}^i n_j$ is the effective block length after $i$ transmissions. In addition, for each block length $n_i$, $i=1, \ldots, K$, we construct a sequence of nested (punctured) polar codes $\{ \Cc(n_i, R_j, \Ac_j^{(i)}) \}_{j=i}^K$ with rates $\{ R_i, \ldots, R_K \}$ such that\footnote{For the ease of exposition, it is assume that every $n_i$ is sufficiently large such that $n_iR_j$ is an integer. Note that using floor function to make every $n_iR_j$ an integer does not change the main results of this paper.}
\begin{equation}
 |\Ac_j^{(i)}| = n_i R_j,    \eqnlabel{SizesOfInformationSets}
\end{equation} for $j \in [K]$. Exact choice of information sets $\Ac_{j}^{(i)}$ is described later in Section~\ref{CodeConstruction}.  We let
\begin{equation}\eqnlabel{Idef}
 {\cal I}^{(i)} \triangleq \bigcup_{j=1}^{i-1} {\cal I}_{i-1}^{(j)}
 \end{equation}
 where $\Ic^{(1)}=[k]$ and ${\cal I}_{i-1}^{(j)}$ is the index set of information bits that are used to convert the polar code in each of the previous transmissions from rate $R_{i-1}$ (that cannot be supported by the channel) to corresponding codes of rate $R_i$, namely, ${\cal I}_{i-1}^{(j)}$ contains the indices of information bits corresponding to $\Ac_{i-1}^{(j)}\setminus\Ac_{i}^{(j)}$.
At transmission $i$, we then use the code $\Cc(n_i, R_i, \Ac_i^{(i)})$ to transmit some part of information bits indexed by ${\cal I}^{(i)}$ as shown in Fig.~\ref{Encoder}. This is possible because ${\cal I}^{(i)}$ and $\Ac_{i}^{(i)}$ are of the same size, that is:
\begin{align}
|{\cal I}^{(i)}| &= \sum_{j=1}^{i-1} |\Ac_{i-1}^{(j)}| - |\Ac_{i}^{(j)}| \stackrel{(a)}{=}\sum_{j=1}^{i-1} q_{i-1}^{(j)}= k - R_i \sum_{j=1}^{i-1} n_j\nonumber\\
 &= k - \frac{k\sum_{j=1}^{i-1} n_j}{\sum_{j=1}^{i} n_j}= n_i R_i \stackrel{(b)}{=} |\Ac_i^{(i)}|,
\end{align}
where $q_{i}^{(j)} \triangleq n_j (R_{i}-R_{i+1})$ for each $i \in [K-1]$, and (a) and (b) follow by \eqnref{SizesOfInformationSets}.

    \begin{figure}[t]
    \centering
   \includegraphics[height=0.8in, width=3in]{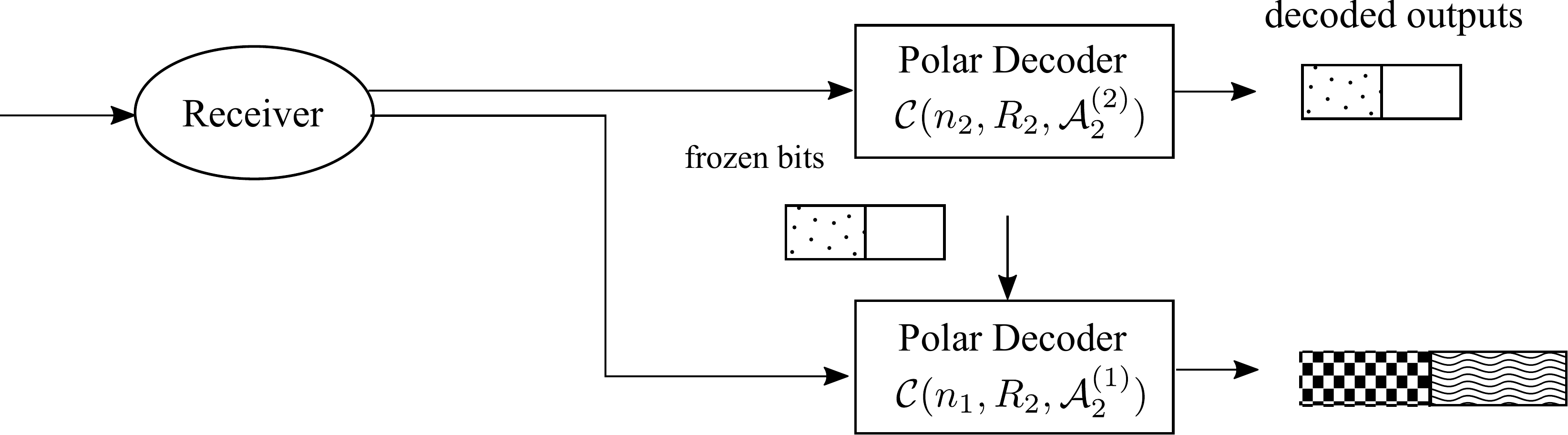}
    \caption{Sequential decoder of rate-compatible polar codes for rate $R_2$.}
         \label{DecoderR2}
    \end{figure}
        \begin{figure}[t]
        \centering
        \includegraphics[height=1.4in, width=3in]{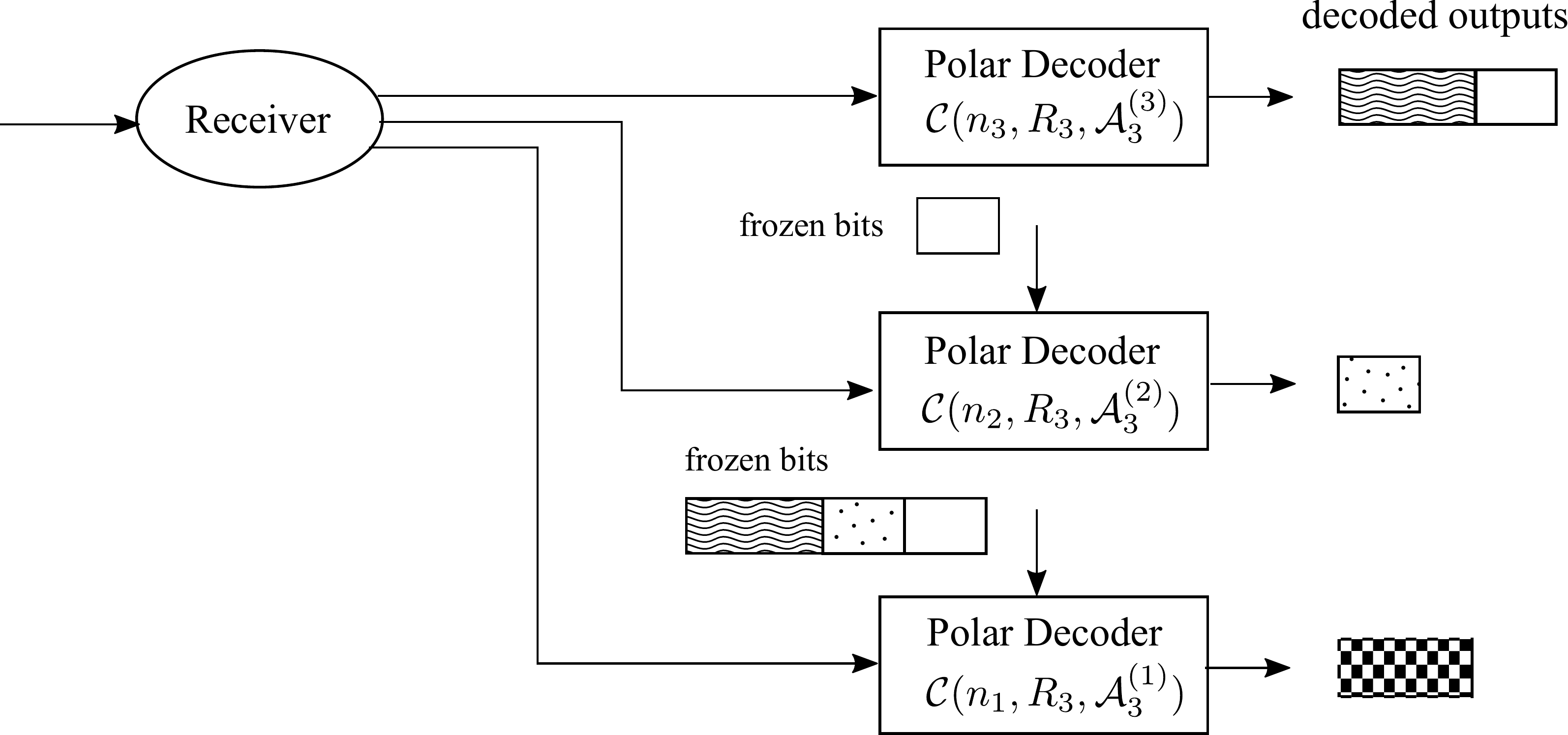}
        \caption{Sequential decoder of rate-compatible polar codes for rate $R_3$.}
             \label{DecoderR3}
        \end{figure}



Code $\Cc (n_i, R_i, \Ac_i^{(i)})$ is the code of the highest rate in the sequence of common block length $n_i$.  Each such sequence satisfies the property \eqnref{degradedset} which will be exploited in decoding. In particular, suppose that $m$ retransmissions, where $m \in [K]$, are needed and hence rate $I(W_m)$ is the highest rate that can be supported by the channel. The decoder starts by decoding the information bits of the polar code $\Cc (n_m, R_m, \Ac_m^{(m)})$.  It then uses some of these decoded bits as frozen bits in the polar code  $\Cc (n_{m-1}, R_{m-1}, \Ac_{m-1}^{(m-1)})$ thereby, due to property \eqnref{degradedset}, turning this code  into a polar code $\Cc (n_{m-1}, R_{m}, \Ac_{m}^{(m-1)})$ in the same sequence (i.e., of the same block length $n_{m-1}$)  but of lower rate $R_m$ which is supported by the channel. Hence, the information bits of this obtained code   can be decoded.  It then repeats this sequential decoding over $m$ stages as shown in Figs. \ref{DecoderR2} and \ref{DecoderR3}, where in each stage, it decodes additional information bits using a polar code of rate $R_m$. As we show later in Theorem~\thmref{capacitytheorem}, the chosen transmit rates $R_i$ as defined in \eqnref{EffectiveRates} will approach the corresponding channel capacity $I(W_i)$, as $n_i$ increases for all $i=1, 2, \ldots, K.$

\subsection{Example}

To explain the main idea in more detail we show an example for $K=3$. In this case, we wish to construct a family of rate-compatible polar codes that supports rates $R_{1}>R_{2}>R_{3}$.

We start by constructing a sequence of nested (punctured) polar codes of block length $n_1$. By Theorem \ref{thm:punc}, we can determine a sufficiently large $n_1$ and associated information sets denoted by ${\cal A}_j^{(1)}$, $j=1,2,3$ of size $|{\cal A}_j^{(1)}| \geq n_1 R_j$ to ensure a tolerable probability of error for each bit-channel in ${\cal A}_j^{(1)}$ for all $j=1,2,3$ when the information sets are applied to channels $W_1$, $W_2$ and $W_3$, respectively.  By Lemma \lemmaref{Korada}, such information sets satisfy the nested property as ${\cal A}^{(1)}_{1} \supseteq {\cal A}^{(1)}_{2} \supseteq {\cal A}^{(1)}_{3}$.  For convenience, we only use subsets of these information sets if necessary such that $|{\cal A}_j^{(1)}| = n_1 R_j$ for all $j=1,2,3$.  The nested subset property can be preserved by first determining ${\cal A}^{(1)}_{3}$ such that $|{\cal A}_3^{(1)}| = n_1 R_3$, then selecting additional bit-channels from ${\cal A}^{(1)}_{2}\setminus {\cal A}^{(1)}_{3}$ to form a new ${\cal A}^{(1)}_{2}$ such that $|{\cal A}_2^{(1)}| = n_1 R_2$, and so forth to form a new ${\cal A}^{(1)}_{1}$ such that $|{\cal A}_1^{(1)}| = n_1 R_1$.  As a result, each code with information set ${\cal A}_j^{(1)}$ enables us to decode $n_1 R_j$ information bits, and we will use these three information sets (i.e., polar codes) to support rates $R_1 > R_2 >R_3$ for the chosen block length $n_1$.

\begin{figure}[t]
 \centering
 \includegraphics[height=2.0in, width=3.2in]{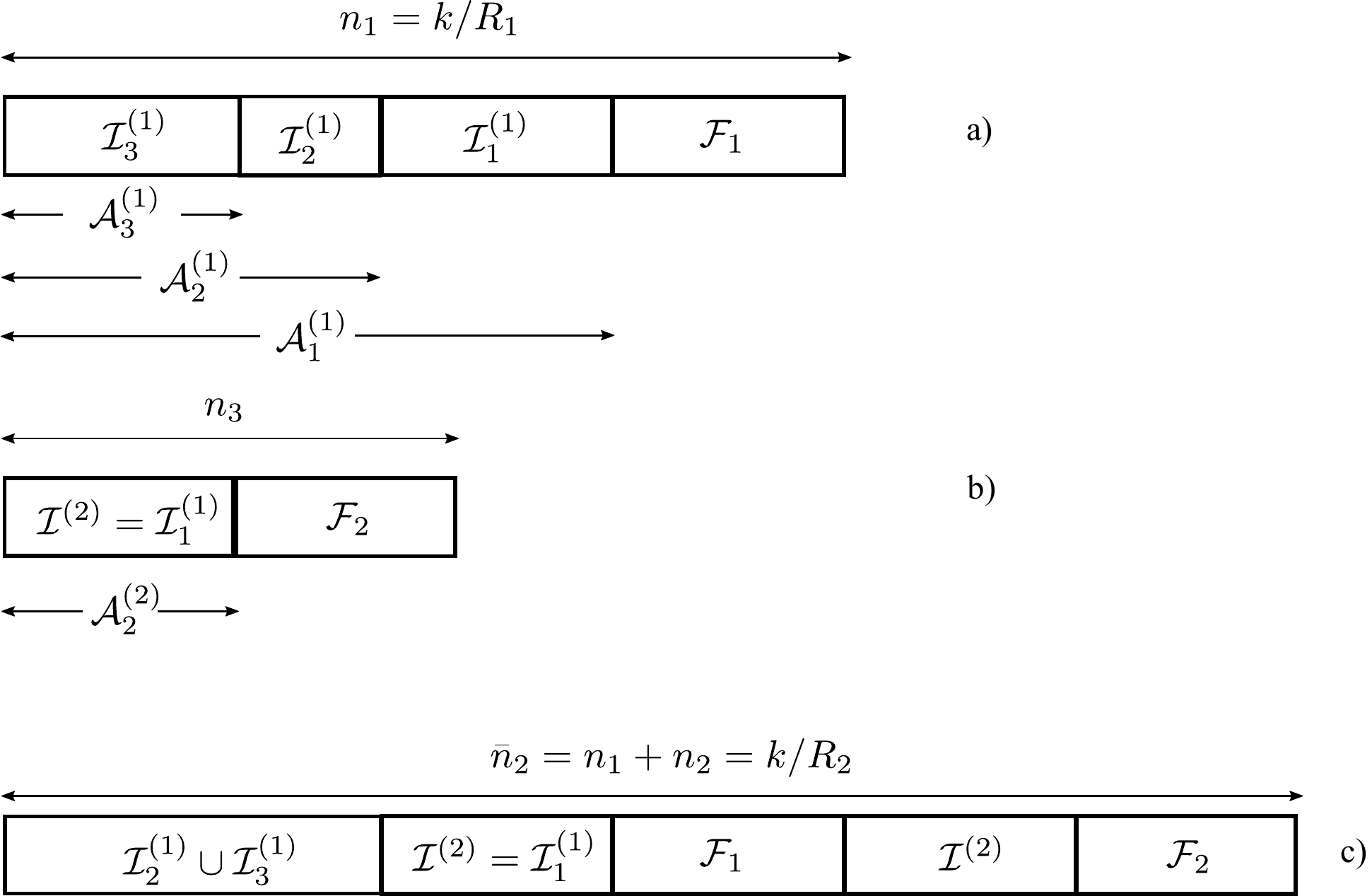}
 \caption{Code construction for $R_2$. ${\cal F}_i$, $i=1,2$ denotes frozen bits. a) Polar codes of rate $R_1$ sent in the first transmission with information bits in ${\cal I}^{(1)} = {\cal I}_{1}^{(1)}\cup {\cal I}_{2}^{(1)} \cup {\cal I}_{3}^{(1)}$; b) Polar code of rate $R_2$ sent in the second transmission with information bits in ${\cal I}^{(2)} = {\cal I}_{2}^{(2)} \cup {\cal I}_{3}^{(2)}$; 1c) Resulting concatenated codeword sent over two transmissions.}
      \label{Fig1}
 \end{figure}

In the first transmission, we use the first (punctured) polar code $\Cc(n_1,R_1,{\cal A}_1^{(1)})$ to transmit $k = |{\cal A}_1^{(1)}|$ information bits at rate $R_1=k/n_1$. Recall that ${\cal I}^{(1)}=\{1,2,\ldots,k\}$. To identify the information bits to be transmitted in the subsequent transmissions, we partition the index set ${\cal I}^{(1)}$ of size $k$ into ${\cal I}^{(1)} ={\cal I}_{1}^{(1)} \cup {\cal I}_{2}^{(1)} \cup {\cal I}_{3}^{(1)}$ such that ${\cal I}_1^{(1)}$, ${\cal I}_{2}^{(1)}$, and ${\cal I}_3^{(1)}$ contain the indices of information bits that will be transmitted through the bit channels in ${\cal A}^{(1)}_{1} \setminus {\cal A}^{(1)}_{2}$, ${\cal A}^{(1)}_{2}\setminus {\cal A}^{(1)}_{3}$, and  ${\cal A}^{(1)}_{3}$, respectively (see Fig.~\ref{Fig1}).  It follows that
\begin{align*}
&|{\cal I}_1^{(1)}| = |{\cal A}^{(1)}_{1}| - |{\cal A}^{(1)}_{2}| = n_1R_1 - n_1R_2\\
&|{\cal I}_2^{(1)}| = |{\cal A}^{(1)}_{2}| - |{\cal A}^{(1)}_{3}| =  n_1 R_2 - n_1 R_3\\
&|{\cal I}_3^{(1)}| = |{\cal A}^{(1)}_{3}| = n_1R_3.
\end{align*}
Sets ${\cal I}_1^{(1)}$ and ${\cal I}_2^{(1)}$ consist of information bits that need to be frozen in respective codes $\Cc (n_1, R_1, {\cal A}_1^{(1)})$ and $\Cc(n_1, R_2, {\cal A}_2^{(1)})$ in order to reduce their rates to $R_2$ and $R_3$, respectively, if subsequent transmissions are needed.
 Intuitively, we expect that the information bits in ${\cal I}_j^{(1)}$ are assigned to better polarized bit channels as $j$ increase.

In the second transmission, we transmit the information bits indexed by ${\cal I}_1^{(1)}$ using a new (punctured) polar code of length $n_2$.  We choose $n_2$ such that $\bar{n}_2 \triangleq n_1+n_2 = k/R_2$ to ensure that, after the second transmission, the effective rate equals $R_2$.  As before, by Theorem \ref{thm:punc}, we can determine a sequence of nested polar codes with information sets denoted by ${\cal A}_j^{(2)}$ of size
$|{\cal A}_j^{(2)}| = n_2 R_j$ for $j=2,3$, and ${\cal A}_2^{(2)} \supseteq {\cal A}_3^{(2)}$. Then, ${\cal I}^{(2)} = {\cal I}_1^{(1)}$  that contains the information bits to be sent in the second transmission. We then use (punctured) polar code $\Cc(n_{2},R_2, {\cal A}_2^{(2)})$ to transmit such information bits, which is possible since
\begin{eqnarray*}
|{\cal I}^{(2)}| = |{\cal I}_1^{(1)}| &=& n_1R_1 - n_1R_2 \\
&=& k - \frac{n_1 k}{n_1+n_2}  = n_2 R_2 = |{\cal A}_2^{(2)}|.
\end{eqnarray*}
As before, for subsequent transmission, we partition ${\cal I}^{(2)}$ into ${\cal I}^{(2)} = {\cal I}_2^{(2)} \cup {\cal I}_3^{(2)}$ such that ${\cal I}_2^{(2)}$ and ${\cal I}_3^{(2)}$ contain respective indices of information bits that are transmitted through the bit channels in  ${\cal A}_2^{(2)} \setminus {\cal A}_3^{(2)}$ and ${\cal A}_3^{(2)}$. Then,
\begin{align*}
  &|{\cal I}_2^{(2)}| = |{\cal A}^{(2)}_{2}| - |{\cal A}^{(1)}_{2}| = n_2 R_2 - n_2 R_3\\
  &|{\cal I}_3^{(2)}| = |{\cal A}^{(2)}_{3}| = n_2 R_3.
\end{align*}
  Finally, for the third transmission, we choose block length $n_3$ such that $\bar{n}_3 = n_1+ n_2 + n_3 = k/R_3$ to attain an effective overall rate of $R_3$ after the third transmission. We encode the information bits indexed by $I^{(3)} = I_{2}^{(1)} \cup I_2^{(2)}$ using a (punctured) polar code with an information set ${\cal A}^{(3)}_{3}$ of size $n_3R_3$, which is possible since
which is possible since
\begin{eqnarray*}
|{\cal I}^{(3)}| = |{\cal I}_{2}^{(1)}| + |{\cal I}_2^{(2)}| &=& (n_1 + n_2)R_2 -(n_1 + n_2)R_3 \\
&=& k -\frac{(n_1 + n_2)k}{n_1 + n_2 + n_3} =  n_3 R_3 = |{\cal A}_3^{(3)}|.
\end{eqnarray*}

\begin{figure}[t]
\centering
\includegraphics[height=1.0in, width=3.4in]{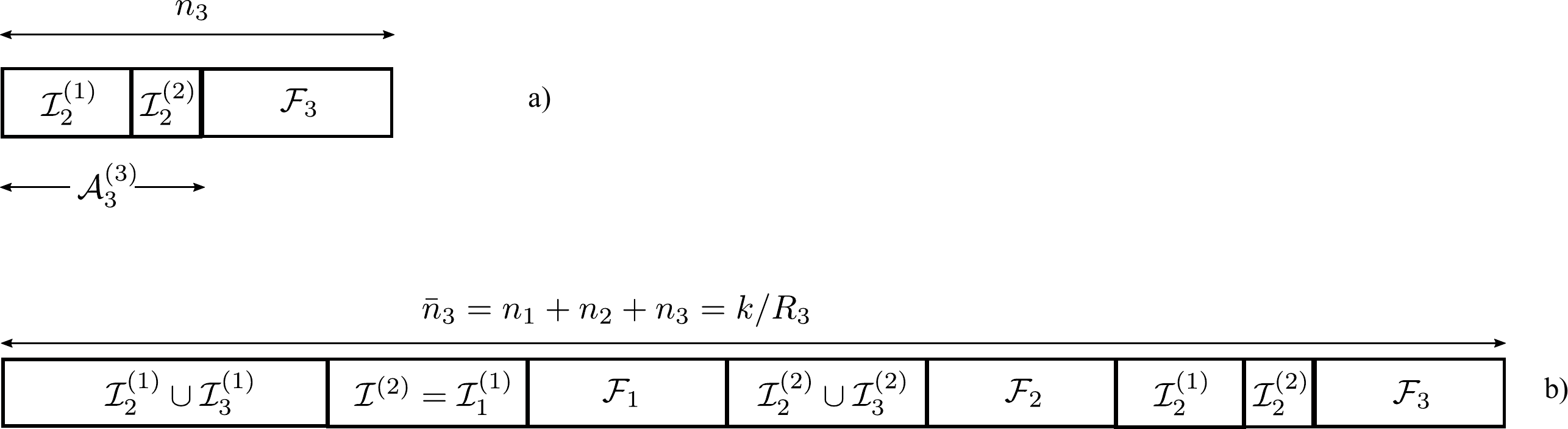}
\caption{Code construction for $R_3$. a) Polar code of rate $R_3$ sent in the third transmission with information bits in ${\cal I}^{(3)} = {\cal I}_{2}^{(1)} \cup {\cal I}_{2}^{(2)}$; b) Resulting concatenated codeword sent over three transmissions.}
     \label{Fig2}
\end{figure}

For $R_2$ and $R_3$, the encoding procedure  and the resulting codeword  are respectively shown in Fig.~\ref{Fig1} and  Fig.~\ref{Fig2}. Note that the obtained codeword of length $k/R_3$ is not a codeword of a polar code. Nonetheless, the decoding procedure will assure that we decode a codeword from a polar code at each rate $R_i, i=1, \ldots ,3$ thereby achieving capacity of the corresponding channel $W_i$.

 In particular, decoder for $\Cc(n_3,R_3,{\cal A}_3^{(3)})$ is first used to decode $R_3n_3$ information bits indexed by ${\cal I}^{(3)}$. Decoded information bits indexed by ${\cal I}_{2}^{(2)}$ are then used as frozen bits in the code $\Cc(n_2, R_2, {\cal A}_2^{(2)})$ in order to produce polar code $\Cc(n_2, R_3, {\cal A}_3^{(2)})$ thereby enabling decoding of $R_3n_2$ information bits indexed by ${\cal I}_3^{(2)}$. We have so far decoded the information bits indexed by ${\cal I}^{(3)} \cup {\cal I}_2^{(1)}$. To decode the rest of information bits, we need to convert code of length $n_1$ and rate $R_1$ that was used for the first transmission, into a code of rate $R_3$. We do so by considering decoded information bits indexed by ${\cal I}^{(3)} \cup {\cal I}_2^{(1)}$ as frozen bits in polar code $\Cc(n_1, R_2, {\cal A}^{(1)}_{1})$ to produce the polar code $\Cc(n_1, R_3, {\cal A}_{3}^{(1)})$ with $R_3 n_1$ information bits indexed by  ${\cal I}_{3}^{(1)}$ that can be decoded. Therefore, we have now decoded all information bits.
\section{General Code Construction} \label{CodeConstruction}

In this section, we describe a general method of constructing rate compatible (punctured) polar codes through  concatenation of generating matrices of multiple polar codes with (possibly) different block lengths.  We refer to the construction process as parallel concatenation and to the resulting class of codes as {\em parallel concatenated polar (PCP) codes}.
\subsection{Parallel Concatenated Polar (PCP) Codes}\label{subsec:PCPC}

We first provide a formal definition of the proposed PCP code:

\begin{definition}\label{def:PCP} Given an information block size $k$ and a set of subblock lengths $\{n_i\}_{i\in [K]}$, a $K$-level PCP code of overall block length $\bar{n}_K$ and rate $R_K$ is characterized by a collection of $K$ sequences of nested polar codes $\{\Cc(n_i,R_j,\Ac_j^{(i)})\}_{j\geq i, i\in [K]}$ and a collection of $K$ bit mappings, $h^{(i)}:[|\Ac_i^{(i)}|] \rightarrow [k]$ (defined below) for  $i\in [K]$, with the following conditions:
\begin{description}
\item[(c.1)] $R_i = k/{\sum_{j=1}^{i} n_j}$ for all $i \in [K]$,
\item[(c.2)] ${\cal A}_i^{(i)} \supseteq {\cal A}_{i+1}^{(i)} \supseteq \cdots \supseteq {\cal A}_{K}^{(i)}$ for each $i\in[K]$,
\item[(c.3)] $|{\cal A}_j^{(i)}| = n_i R_j$ for all $j \geq i$ and $i \in [K]$.
\end{description}Here, (c.1) is required for any rate-compatible code as given in Lemma~\lemmaref{RateLengthRelation}, and (c.2) and (c.3) are additional conditions required  for the construction of the rate-compatible polar code.
\end{definition}

Note that a total of $K(K+1)/2$ (punctured) polar codes, covering different information-set sizes $|\Ac_j^{(i)}|$ for all $j \geq i$ and $i \in [K]$, is involved in a $K$-level PCP code. The $k \times R_K$ generator matrix of a $K$-level PCP code is fully determined by $( \{\Cc(n_i,R_j,\Ac_j^{(i)}\}_{j\geq i, i\in [K]}, \{h^{(i)}\}_{i\in [K]} )$ and is a concatenation of submatrices of $\{\Pv_{n_i}\}_{i\in[K]}$ in the form
\begin{equation}
\Gv_K =
\begin{bmatrix}
\Sv_1 & \Sv_2 & \ldots & \Sv_K
\end{bmatrix},
\end{equation}
where $\Sv_i$ is a $k \times n_i$ matrix whose non-zero rows come from the rows of $\Pv_{n_i}$ and are indexed by the set ${\cal I}^{(i)} = h^{(i)}([|\Ac_i^{(i)}|])$.  More precisely, the $m$-th row of the matrix   $\Pv_{n_i}^{(\Ac_i^{(i)},\Ac_{i+1}^{(i)},\cdots,\Ac_{K}^{(i)})}$, which is a row-permuted submatrix of $\Pv_{n_i}$, is placed at the $h^{(i)}(m)$-th row of $\Sv_i$, for every $m \in [|\Ac_i^{(i)}|]$, while all other $(k - n_i R_i)$ rows of $\Sv_i$ are zero, where $|\Ac_i^{(i)}| = n_i R_i$.  The matrix  $\Sv_i$ defines the "$i$-th level" of the PCP code.

A key feature of a PCP code is that it is sequentially decodable level-by-level if the bit-mapping functions $\{h^{(i)}(\cdot)\}_{i\in[K]}$ are properly related in accordance with $\{\Ac_j^{(i)}\}$.   We define their proper relationships in a recursive manner in the following.  Suppose that we are given the matrices $\{ \Pv_{n_j}^{(\Ac_{j}^{(j)},\Ac_{j+1}^{(j)},\cdots,\Ac_{K}^{(j)})} \}_{j\in[K]}$ as defined in \eqnref{StackedPolarGenMatrix} and that we have already determined the submatrices $\{ \Sv_j \}_{j=1}^i$ of $\Gv_K$ up to the $i$-th level for some $i \ge 1$. Also suppose that we know the corresponding bit mapping $h^{(j)}: [n_jR_j] \rightarrow [k]$ that specifies the non-zero rows of $\Sv_j$  that contains the rows of $\Pv_{n_j}^{(\Ac_j^{(j)},\Ac_{j+1}^{(j)},\cdots,\Ac_{K}^{(j)})}$ for all $j \in [i]$, with $h^{(1)}(\cdot)$ simply defined as $h^{(1)}(m)=m$ for $m \in [k]$. We now define $h^{(i+1)}(\cdot)$ in terms of $\{h^{(m)}(\cdot)\}_{m\in[i]}$.
To simplify notation, we define $Q_0^{(i+1)} \triangleq 0$ and for each $j \in [i]$,
\begin{equation}
Q_{j}^{(i+1)} = \sum_{l=1}^{j} q_i^{(l)}.
\end{equation}
Now we partition the domain of $h^{(i+1)}(\cdot)$, namely $[n_{i+1} R_{i+1}]$ into disjoint sets as
\begin{equation}
[n_{i+1}R_{i+1}] = \bigcup_{j=1}^i {\cal J}_j^{(i+1)},
\end{equation}
where
\begin{equation*}
{\cal J}_j^{(i+1)} \triangleq \left\{ Q_{j-1}^{(i+1)}+1, Q_{j-1}^{(i+1)}+2, \cdots , Q_{j-1}^{(i+1)}+ q_i^{(j)} \right\},
\end{equation*}
is a set of consecutive integers for all $j \in [i]$. We can now define an bit mapping $h^{(i+1)}: [n_{i+1}R_{i+1}] \rightarrow {\cal I}^{(i+1)} \subseteq [k]$  for $S_{i+1}$ in a piece-wise fashion as
\begin{equation*}
h^{(i+1)}(m) = h^{(j_m)} \left( m - Q_{j_m-1}^{(i+1)} +n_{j_m}R_{j_m} - \sum_{l=j_m}^{i+1} q_{l}^{(j_m)} \right),
\end{equation*}
for $m\in [n_{i+1}R_{i+1}]$, where $j_m$ denotes the index of the interval ${\cal I}_{j_m}^{(i+1)}$ that contains integer $m$.
Note that the image ${\cal I}_i^{(j)} \triangleq h^{(i+1)}({\cal J}_j^{(i+1)})$ on the set ${\cal J}_j^{(i+1)}$ is the index set of information bits that needs to be decoded, and subsequently frozen, in order to convert the polar code $\Cc(n_j,R_i,\Ac_i^{(j)})$ of rate $R_i$ to the polar code $\Cc(n_j,R_{i+1},\Ac_{i+1}^{(j)})$ for all $j\in[i]$ and $i\in [K-1]$ within the respective nested polar code sequences.  With this definition of $\{h^{(i)}(\cdot)\}_{i\in[K]}$, it can be easily shown  that a $K$-level PCP code can be decoded sequentially from the polar code at level $K$ back to level 1.

A $K$-level PCP code $( \{\Cc(n_i,R_j,\Ac_j^{(i)}\}_{j\geq i, i\in [K]}, \{h^{(i)}\}_{i\in [K]})$ clearly induces a family of rate-compatible linear codes ${\bf\cal C}=\{\Cc_{1}^{\bar{n}_1}, \Cc_{2}^{\bar{n}_2}, \cdots, \Cc_{K}^{\bar{n}_K}\}$ with each member $\Cc_{i}^{\bar{n}_i}$ having block length $\bar{n}_i \triangleq \sum_{j \in [i]} n_j$, rate $R_i$, and a generator matrix  $\Gv_i =
	\begin{bmatrix}
	\Sv_1 & \Sv_2 & \cdots & \Sv_i
	\end{bmatrix}$,
since $\{\Gv_i\}_{i\in[K]}$ forms a sequence of nested generating matrices, i.e. $\Gv_i \subseteq \Gv_{i+1}$ for all $i\in[K-1]$.  The $K$-level PCP code with generator matrix $\Gv_K$ may be viewed as the "mother" code of the lowest rate in the rate compatible family.

\subsection{Sequential Decoding} \label{SequentialDecoder}

\vspace{0.15cm}
To decode the constructed  parallel concatenated  polar code we use sequential decoder shown in Algorithm 1.
\begin{algorithm}[h]
\caption{Decoding algorithm}\label{decoder}
\begin{algorithmic}[1]
\Procedure{Decoder}{$\yv^{\bar n_K}$}\Comment{Input: received vector $\yv^{\bar n_K}$}
\State Decode $n_KR_K$  bits $\Ic^{(K)}$ using $\Cc(n_K, R_K, \Ac_K^{(K)})$
\For{$i = K-1, \ldots, 1$}
\State Use $\bigcup_{j=i}^{K-1} \Ic_j^{(i)}$ as frozen bits in $\Cc(n_{i}, R_{i}, \Ac_{i}^{(i)})$ to get $\Cc(n_{i}, R_K, \Ac_K^{(i)})$
\State Decode $n_{i}R_K$  bits $\Ic_K^{(i)}$ using $\Cc(n_{i}, R_K, \Ac_K^{(i)})$
\EndFor\label{euclidendwhile}

\State \textbf{return} $\Ic^{(1)} = \bigcup_{i=1}^K \Ic_K^{(i)}$\Comment{Output: decoded bits $\Ic^{(1)}$}
\EndProcedure
\end{algorithmic}
\end{algorithm}
In the  decoding given by Algorithm 1, information bits are decoded by $K$ polar decoders sequentially and we use the fact that $\Ic^{(i)}$ for $i=1, \ldots, K$ are related by \eqnref{Idef}. The number of bit decisions made by the sequential decoder is independent of the number of decoding stages $K$ and it equals  the number of information bits $k$. The sequential decoder approach is optimal in the sense that it achieves the capacity as all block lengths approach infinity.


\section{Simulation Results: Finite Length} \label{SimulationResults}
\begin{figure}[t]
 \centering
  \includegraphics[height=2.6in, width=3.5in]{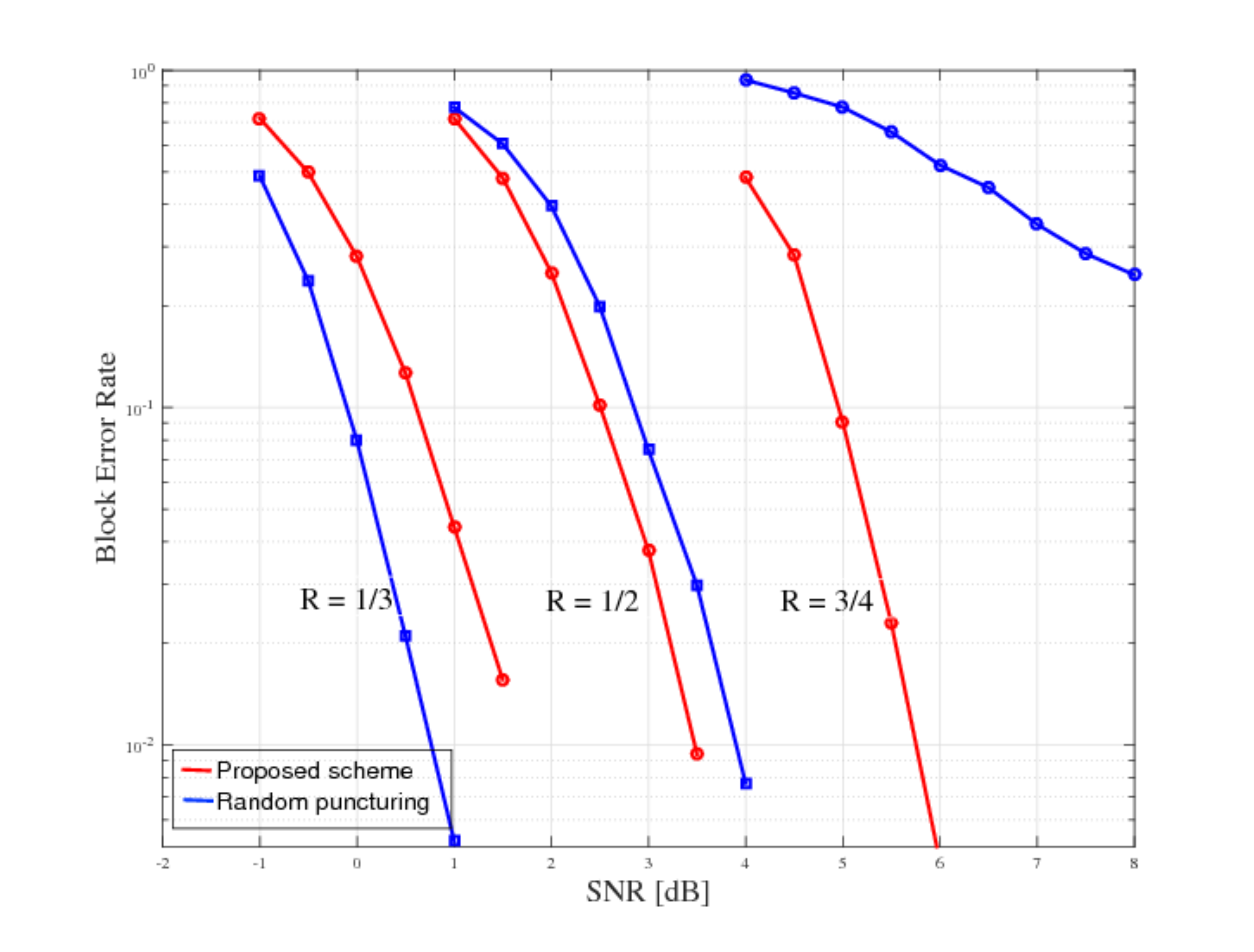}
 \caption{Performance of the proposed scheme vs. puncturing.}
      \label{FigPerformance}
 \end{figure}

Performance comparison of the proposed scheme with random puncturing is shown in Fig.~\ref{FigPerformance}.  In Fig.~\ref{FigPerformance}, the mother code in our scheme has rate $R=3/4$ and the codes for lower rates $R=1/2$ and $R=1/3$  are obtained using parallel concatenated polar codes as described in previous sections (see Table~\ref{table} for specific polar code descriptions). Since $n_{3}$ is not of the form of powers of two, a punctured polar code was used for that block length where the locations of punctured bits are uniformly distributed and the information set is optimized by taking  puncturing into account.
In contrast,  for random puncturing approach, the information set is optimized for the mother code with rate $R=1/3$ and higher rates $R=1/2$ and $R=3/4$ are obtained by random puncturing. As a mother code, we used a polar code with block length $512$ and information bits $171$. We observe that  our scheme significantly outperforms random puncturing at  rate $R=3/4$ where a large number of bits need to be punctured in the latter scheme. At rate $R=1/3$, random puncturing outperforms parallel concatenated polar codes. This may be  due to the use of suboptimal sequential decoder where an error probability is computed as $1-P(\{\mbox{all component decoders have no error}\})$. Namely, in the sequential decoder, it is assumed that retransmitted information bits are decoded based only on current reception. However, reception from the previous retransmissions still contain useful information, and thus finite-block length performance can be improved by employing a soft decoder; each component code receives a soft information from other component codes and exploits it as a priori information as performed in Turbo code \cite{Berrou}. The study of an enhanced decoder is left for a future work.

\begin{table}[h!]
\begin{center}
\caption{Code Constructions}
\label{table}
\begin{tabular}{|c|c|c|c|}
  \hline
   &  $R_{1}=3/4$&$R_{2}=1/2$  &$R_{3}=1/3$  \\
   \hline
   $n_{1}=256$ & $k=192$  & $k=128$ & $k=85$ \\
   \hline
   $n_{2}=128$&  & $k=64$ & $k=42$\\
   \hline
   $n_{3}=195$ &  &  & $k=65$ \\
  \hline
\end{tabular}
\end{center}
\end{table}


\section{Proofs} \label{Proofs}
In this section, we provide the proofs of our main theorems.

\subsection{Proof of Lemma \lemmaref{RateLengthRelation}} \label{proof:lemma1}
\begin{IEEEproof}
Since a rate-compatible code family is designed for a fixed number of information bits, we have, from the definition of the effective rate after each transmission,
\begin{equation}
k = n_1 R_1 = \bar{n}_i R_i =  R_i \sum_{j=1}^{i} n_j   \eqnlabel{SameNumberInfoBits}
\end{equation}
for all $i \in \{2,3,\cdots,K\}$.  Condition (a) in the statement of the lemma follows immediately from \eqnref{SameNumberInfoBits}.   Condition (b) can be readily verified by substituting Condition (a) into the right side of Condition (b).  Finally, by summing Condition (b) for consecutive values of index $i$, \eqnref{SameNumberInfoBits} can be obtained, which in turn implies Condition (a).  Conditions (a) and (b) are thus equivalent.
\end{IEEEproof}

\subsection{Proof of Theorem \ref{thm:punc}}\label{subsec:thm1proof}

We first prove the following Proposition~\ref{thm:punc_epsilon} and then apply it with a slightly larger target puncturing fraction $\alpha' = \alpha/(1-\epsilon)$, $\epsilon = \frac{1-\alpha}{1+2 \alpha R /(I(W)-R)} = \frac{1-\alpha'}{\alpha'} \frac{I(W)/R-1}{2} > 0$, and a slightly higher target rate $R' = R (1+\epsilon \frac{\alpha'}{1-\alpha'}) = \left(I(W)+R\right)/2 < I(W) $.  We obtain $m^* \in \mathbb{N}$ such that for each $m \geq m^*$, we have a punctured polar code with block length $n=\lfloor(1-(1-\epsilon)\alpha')2^m \rfloor =\lfloor (1-\alpha) 2^m \rfloor$, an information set $|{\cal A}_{m}| \geq  2^m (1-\alpha')R' = 2^m (1-(1-\epsilon)\alpha') R = 2^m (1-\alpha) R \geq n R$, and
	\begin{equation}
	P_{e,j,m} \leq  C(\epsilon,\alpha') \left( 2^{-2^{m\beta}}\right),
	\end{equation} for all $j\in {\cal A}_{n}$ and
	for some constant $C(\epsilon,\alpha')$.  By simply picking a large enough constant $C(\epsilon,\alpha')$ such that $C(\epsilon,\alpha') 2^{-2^{m^*\beta}} \geq 1$, the desired result \eqnref{BitErrorCondition} follows for all $m \in \mathbb{N}$.
%
\begin{MyProposition}\label{thm:punc_epsilon} Consider any B-DMC $W$ with $I(W) > 0$. For any fixed $R < I(W)$, $\beta < \frac{1}{2}$, $0 < \epsilon < 1$, and $0<\alpha<1$, there exists a sequence of punctured polar codes, each with block length $n = \lfloor(1-(1-\epsilon)\alpha) 2^m \rfloor$ and associated information set ${\cal A}_{m} \subset [2^m]$, for all integer $m \geq m^{\star}(\epsilon,\alpha)$, such that $|{\cal A}_{m}| \geq 2^m (1-\alpha)R$ and
	\begin{equation}\eqnlabel{BitErrorCondition2}
	P_{e,j,m} \leq	O\left(2^{-2^{m\beta}}\right) = O\left(2^{-n^{\beta}}\right),
	\end{equation} for all $j\in {\cal A}_{m}$.
\end{MyProposition}

\begin{IEEEproof}
 We fix $\alpha$, $\beta$, $\epsilon$, and $R$. For the proof, we show the existence of a sequence of capacity-achieving punctured polar codes using a random puncturing argument. For a given target puncturing fraction $\alpha$ and block length $n_{u}$, an effective puncturing fraction $\alpha_{n_{u}}$ is computed as
 \begin{equation}
 \alpha_{n_{u}} = \frac{n_{u}-\lfloor(1-(1-\epsilon)\alpha) n_{u} \rfloor}{n_{u}},
 \end{equation} where notice that the effective puncturing fraction depends on the block length $n_{u}$. Since $\alpha_{n_{u}} \leq (1-\epsilon)\alpha + \frac{1}{n_{u}}$, there exist $\epsilon' < \epsilon$ and $n_{u}(\epsilon')$ such that
 \begin{equation}\eqnlabel{alpha_bound}
 \alpha_{n_{u}} \leq (1-\epsilon')\alpha,
 \end{equation} for all $n_{u} \geq n_{u}(\epsilon')$.

Let ${\cal P}(n_u, \alpha_{n_{u}})$ denote a set of possible puncturing patterns in $\{0,1\}^{n_u}$ with (exact) puncturing fraction $\alpha_{n_{u}}$. Random puncturing implies that a puncturing pattern $p^{n_u}$ is chosen from $p^{n_u} \in {\cal P}(n_u,\alpha_{n_{u}})$ with probability $f(p^{n_u})$. Then, a random puncturing is defined by specifying a puncturing set ${\cal P}(n_u,\alpha_{n_{u}})$ and a puncturing distribution $f(p^{n_u})$. In the proof, we will design a specific puncturing set  ${\cal P}(n_u, \alpha_{n_{u}})$ and define  distribution $f(p^{n_u})$.  We let $ \bar P_{e,j,m}$  denote the {\em average} bit error probability of $j$-th polarized bit channel followed by random puncturing, where the average is performed with respect to puncturing distribution $f(p^{n_u})$. We will show that there exist an information set $A_{m}$, a puncturing set ${\cal P}(n_u,\alpha_{n_{u}})$, and a puncturing distribution $f(p^{n_u})$, such that
\begin{equation}
\bar P_{e,j,m} \leq O\left(2^{-2^{m\beta}}\right),
\end{equation} for all $j \in A_{m}$. Therefore, there exists at least one puncturing pattern $\hat{p}^{n_{u}} \in {\cal P}(n_u,\alpha_{n_{u}})$ that satisfies \eqnref{BitErrorCondition2}.
	
	First, we find an information set for a punctured polar code as follows. Given the {\em target} puncturing fraction $\alpha$, we form the {\em cascade} channel of binary erasure channel with erasure probability $\alpha$ (denoted by BEC($\alpha$)) and $W$, which is denoted by $W(\alpha)$. Obviously, the resulting channel $W(\alpha)$ is B-DMC. From \cite[Theorem 2]{Arikan2009}, for $W(\alpha)$ and fixed $R<I(W)$ (i.e., $R'=(1-\alpha)R < I(W(\alpha))$), there exists a sequences of ${\cal A}_{m} \subset [2^{m}]$ such that
\begin{align}
|{\cal A}_{m}| &\geq 2^{m}R'=2^{m}(1-\alpha)R\\
Z(W_{n_{u}}^{(i)}(\alpha)) &\leq O(2^{-5m/4})
\end{align} for all $i \in {\cal A}_{m}$. We use the stronger result on the rate polarization from \cite{TelatarArikan2009} that $Z(W_{n_{u}}^{(i)}(\alpha)) \leq O(2^{-2^{m\beta}})$. Notice that this information set will be used for the punctured polar code.
	
	Next, we show that there exist a puncturing set ${\cal P}(n_u,\alpha_{n_u})$ and a puncturing distribution $f(p^{n_u})$ such that
$\bar P_{e,j,m} \leq O(2^{-2^{m\beta}})$ for all $j \in {\cal A}_{m}$. Let $y_{e}^{n_u}$ and $y^{n_u}$ denote the channel outputs of $W(\alpha)$ and $W$, respectively. Recall that $Z(W_{n_u}^{i}(\alpha))$ denotes the upper bound on the error probability of polarized channel $j$ (of the cascade channel), defined by
\begin{align}
P({\cal E}_{j}) &\triangleq \sum_{p^{n_{u}}\in \Omega_{n_{u},\alpha}}P({\cal E}_{j,p^{n_{u}}})P(p^{n_{u}}),\label{eq:defE}
\end{align}where ${\cal E}_{j,p^{n_{u}}}$ denotes a {\em conditional} error event of SC decoder for the given erasure pattern $p^{n_{u}}$ (induced by BEC($\alpha$)) and $\Omega_{n_{u},\alpha}$ denotes the set of all possible erasure patterns induced by BEC($\alpha$).

	Define the {\em empirical} probability mass function (pmf) of $p^{n_u}$ as
	\begin{equation*}
	\pi(x|p^{n_u}) = \frac{|\{i:p_{i}=x\}|}{n_u} \mbox{ for } x \in \{0,1\},
	\end{equation*} where ``$0$" represents an erasure. Let $X_{1},X_{2},...$ be a sequence of independent and identically distributed random variables $X_{i} \sim P(x_{i}=0) = \alpha$. For $X \sim f(x)$ and fixed $\epsilon' \in (0,1)$, define the set of $\epsilon'$-typical $n$-sequences $p^{n_u}$ as
	\begin{equation*}
	{\cal T}_{\epsilon'}^{(n_{u})}(\alpha) =\{p^{n_u}: |\pi(x|p^{n_u}) - f(x)| < \epsilon' f(x),
	\end{equation*} for all $x \in \{0,1\}$.	
	Clearly, we have that ${\cal T}_{\epsilon'}^{(n_{u})}(\alpha) \subseteq \Omega_{n_{u},\alpha}$. Further, by the Law of Large Numbers, there exists $n'_{u}(\epsilon')$ such that $P({\cal T}_{\epsilon'}^{(n_{u})}(\alpha)) > 1-\epsilon'$ for all $n_{u} \geq n'_{u}(\epsilon')$. We let $2^{m^{\star}(\epsilon,\alpha)} = \max\{n_{u}(\epsilon'),n'_{u}(\epsilon')\}$. From now on, it is assumed that $n_{u} \geq \max\{n_{u}(\epsilon'),n'_{u}(\epsilon')\}$.
	In (\ref{eq:defE}), by replacing $\Omega_{n_{u},\alpha}$ by ${\cal T}_{\epsilon'}^{(n_{u})}(\alpha)$,  we obtain that
	\begin{align}
	P({\cal E}_{j}') &\triangleq \sum_{p^{n_u} \in {\cal T}_{\epsilon'}^{(n_{u})}(\alpha)} P({\cal E}_{j,p^{n_{u}}}) \frac{P(p^{n_u})}{1-\epsilon'+\delta}\nonumber\\
	&\leq \frac{1}{1-\epsilon'} P({\cal E}_{j}),\label{eq:LB}
	\end{align}where we chose $\delta > 0$ such that $\sum_{p^{n_u} \in {\cal T}_{\epsilon'}^{(n_{u})}(\alpha)} \frac{P(p^{n_u})}{1-\epsilon'+\delta} = 1$. Notice that such $\delta$ always exists since $P({\cal T}_{\epsilon'}^{(n_{u})}(\alpha)) > 1-\epsilon'$. Notice that, for any $p^{n_u} \in {\cal T}_{\epsilon'}^{(n_{u})}(\alpha)$, we have that  $|\pi(0|p^{n_u}) - \alpha| < \epsilon'\alpha$ and $\alpha_{n_{u}}\leq(1-\epsilon')\alpha$ since $n_{u} \geq n_{u}(\epsilon')$. For any $p^{n_u} \in {\cal T}_{\epsilon'}^{(n_{u})}(\alpha)$ define a mapping $\xi: {\cal T}_{\epsilon'}^{(n_{u})}(\alpha) \rightarrow {\cal P}(n_u,\alpha_{n_{u}})$ such that $\hat{p}^{n_u} = \xi(p^{n_u})$ is obtained by removing erasures in $p^{n_u}$ until $\pi(0|\hat{p}^{n_u}) = \alpha_{n_{u}}$, and the positions of erasures removed are chosen with natural ordering. Using this, define
	\begin{equation}
	P(\hat{p}_{n_{u}}) \triangleq \sum_{p^{n_u}: \xi(p^{n_u}) = \hat{p}^{n_u}}\frac{P(p^{n_u})}{1-\epsilon'+\delta}.\label{eq:dist}
	\end{equation} Clearly, $\sum_{\hat{p}^{n_u} \in \xi\left({\cal T}_{\epsilon'}^{(n_{u})}(\alpha)\right)} P(\hat{p}^{n_u}) = 1$. Then, we show that
	\begin{align}
	&P({\cal E}_{j}'') \triangleq \sum_{\hat{p}^{n_u} \in \xi\left({\cal T}_{\epsilon'}^{(n_{u})}(\alpha)\right)} P({\cal E}_{j,\hat{p}^{n_{u}}})P(\hat{p}^{n_u})\nonumber\\
	&= \sum_{\hat{p}^{n_u} \in \xi\left({\cal T}_{\epsilon'}^{(n_{u})}(\alpha)\right)} P({\cal E}_{j,\hat{p}^{n_{u}}}) \sum_{p^{n_u}: \xi(p^{n_u}) = \hat{p}^{n_u}} \frac{P(p^{n_u})}{1-\epsilon'+\delta}\nonumber\\
	&\stackrel{(a)}{\leq} \sum_{\hat{p}^{n_u} \in \xi\left({\cal T}_{\epsilon'}^{(n_{u})}(\alpha)\right)} \sum_{p^{n_u}: \xi(p^{n_u}) = \hat{p}^{n_u}} P({\cal E}_{j,p^{n_{u}}}) \frac{P(p^{n_u})}{1-\epsilon'+\delta}\nonumber\\
    &= P({\cal E}_{j}'), \label{eq:UB}
	\end{align} where (a) is due to the fact that $P({\cal E}_{j,\hat{p}^{n_{u}}})\leq P({\cal E}_{j,p^{n_{u}}})$ for any $p^{n_{u}} \in \xi^{-1}(\{\hat{p}^{n_{u}}\})$, i.e., deleting erasures does not increase an error. From (\ref{eq:LB}) and (\ref{eq:UB}), we obtain that
	\begin{equation}
	P({\cal E}_{j}'') \leq \frac{1}{1-\epsilon'} P({\cal E}_{j}).\label{eq:bound}
	\end{equation}
	
	We are now ready to define our random puncturing with puncturing fraction $\alpha_{n_{u}}$ by choosing
	\begin{align}
	{\cal P}(n_u,\alpha_{n_{u}}) &= \xi\left({\cal T}_{\epsilon}^{(n)}(\alpha)\right)\label{eq:Pset}\\
	f(p^{n_u})&=P(\hat{p}^{n_u}),\label{eq:Pdist}
	\end{align}
	where $P(\hat{p}^{n_u})$ is defined in (\ref{eq:dist}). From (\ref{eq:bound}), we can show that, for all $n_{u} \geq \max\{n_{u}(\epsilon'),n'_{u}(\epsilon')\}$ (equivalently, $m \geq m^{\star}(\epsilon,\alpha)$),
	\begin{equation*}
	\bar P_{e,j,m}=P({\cal E}_{j}'')  \leq \frac{1}{1-\epsilon'}P({\cal E}_{j}) \leq Z(W(\alpha)_{n_u}^{(j)}) \leq O\left(2^{-2^{m\beta}}\right),
	\end{equation*} for all $j \in {\cal A}_{m}$ where ${\cal A}_{m}$ is defined over the cascade channel $W(\alpha)$.  Therefore, there exists at least one puncturing pattern $\hat{p}^{n_{u}} \in {\cal P}(n_u,\alpha_{n_{u}})$ that satisfies \eqnref{BitErrorCondition2}. This completes the proof.

\end{IEEEproof}

\subsection{Proof of Theorem \thmref{capacitytheorem}}\label{subsec:thm2proof}

For  given degraded channels $W_1 \succeq W_2 \succeq \ldots \succeq W_K$, we consider a family of rate-compatible polar code $(\Cc_{1}^{\bar{n}_{1}}, \ldots \Cc_K^{\bar{n}_K})$ with respective block lengths ${\bar n_1}< \ldots < {\bar n_K}$ and corresponding rates $R_1 > \ldots > R_K$ given by  \eqnref{EffectiveRates}.

As described in Section~\ref{subsec:PCPC}, this code is fully characterized by a collection of $K$ sequences of nested (punctured) polar codes $\{\Cc(n_i,R_{j},A_j^{(i)})\}_{j\geq i, i\in [K]}$ which satisfy the three conditions in Definition~\ref{def:PCP}  as
\begin{description}
\item[(c.1)] $R_i = k/\bar{n}_{i}$ for all $i \in [K]$,
\item[(c.2)] ${\cal A}_i^{(i)} \supseteq {\cal A}_{i+1}^{(i)} \supseteq \cdots \supseteq {\cal A}_{K}^{(i)}$ for each $i\in[K]$,
\item[(c.3)] $|{\cal A}_j^{(i)}| = n_i R_j$ for all $j \geq i$ and $i \in [K]$.
\end{description}
Note that the number of information bits equals $|A_{1}^{(1)}|$.

Fix $R_{i}$, $i=1,\ldots,K$, such that $I(W_{i}) > R_{i}  > I(W_{i}) - \epsilon$ for any $\epsilon >0$. We first explain how to choose block lengths $\{n_{i}\}_{i \in [K]}$ and information sets $\{{\cal A}_{j}^{(i)}\}_{j \geq i, i \in [k]}$, in order to construct the above polar codes. We choose $n_1 = (1-\alpha_1) 2^{m_1}$ for some $\alpha_1 \in [0,1)$ and $m_1 \in \mathbb{N}$. Furthermore, we  choose, for $i=2, \ldots, K$ lengths of the form $n_i = (1-\alpha_i) 2^{m_i}$ where $\alpha_i \in [0,1)$ and $m_i \in \mathbb{N}$ as
\begin{equation}\eqnlabel{l}
n_{i} = R_{1}\left(\frac{1}{R_{i}} - \frac{1}{R_{i-1}} \right)n_{1}.
\end{equation}  Notice that there exist $m_{1}^{*}$ and $\alpha_{1}$ such that all $n_{i}$ is an integer for all $i=2,\ldots,K$, since each $R_{i}$ is a rational number. Assuming that $m_{1} \geq m_{1}^{*}$, the $n_{i}$ can be expressed in the form of $n_{i} = (1-\alpha_{i})2^{m_{1} + l_{i}}$ for some non-negative integer $l_{i}$, i.e., $m_{i} = m_{1} + l_{i}$ for all $i=2,\ldots,K$.

%
%

From Theorem \ref{thm:punc}, for fixed $n_{i}$ and channels $\{W_{j}\}_{j=i,\ldots,K}$, there exist information sets $\{{\cal A}_{j}^{(i)}\}_{j=i,\ldots,K}$ such that
\begin{equation}\eqnlabel{Adef}
|{\cal A}_{j}^{(i)}| = R_{j} n_{i},
\end{equation} and an error probability $P_{e,l,m_i}$ is bounded by $O(2^{- 2^{m_{i}\beta}})$ for any fixed $\beta < \frac{1}{2}$ and $l \in \Ac_j^{(i)}$. Since $m_{1} \leq m_{i}$ for all $i\geq 2$, the error probability is bounded by $O(2^{- 2^{m_{1}\beta}})$ for all punctured polar codes that used in sequential decoders. Therefore, there exists a sufficiently large $m_{1}^{\dag}\geq m_{1}^{*}$ such that for all $m_{1} \geq m_{1}^{\dag}$, the error probability condition in \eqnref{ProbCloseToZero} holds.

Finally, we show that our choices of $\{n_{i}\}_{i \in [K]}$ and $\{A_{j}^{(i)}\}_{j\geq i, i \in [K]}$ satisfy the three conditions that are required to hold by our code construction. From \cite[Lemma 4.7]{KoradaPhD2009}, it follows that $A_{i}^{(i)} \supseteq A_{i+1}^{(i)} \supseteq \ldots \supseteq A_{K}^{(i)}$ for each $i \in [K]$. The condition (c.3) immediately holds from \eqnref{Adef}. We then show that the condition (c.1) is satisfied since for all $i \in [K]$, we have that
\begin{align}
\bar{n}_{i} = \sum_{j=1}^{i} n_{j} \stackrel{(a)}{=} \frac{R_{1}}{R_{i}}n_{1}=\frac{k}{R_i},
\end{align}where $(a)$ follows by \eqnref{l} and by telescoping sum. This completes the proof.

\subsection{Proof of Theorem \thmref{theoremnopuncturing}} \label{subsec:thm3proof}

We first prove the sufficiency of \eqnref{Iassumption}. Fix $\ell_{i} \in \mathbb{Z}$ for $i=1,\ldots, K$. Then, it is assumed that
\begin{equation}\eqnlabel{Iassumption1}
I(W_i) = \frac{I(W_1)}{1 + \sum_{j=2}^i 2^{\ell_j}},
\end{equation} for $i=1,\ldots,K$. The proof follows from the proof of Theorem~\thmref{capacitytheorem} with the additional constraints that, due to the length limitation of polar codes, $n_{i}$ should have the form of powers of two for all $i \in [K]$.

Fix $R_{1}$ such that $ I(W_{1}) > R_{1} > I(W_{1}) - \epsilon$ for any $\epsilon >0$. Then, we choose $R_{i}$, $i=2,\ldots,K$, such that
\begin{equation}
 R_i = \frac{R_1}{1 + \sum_{j=2}^i 2^{l_{j}}}.   \eqnlabel{RateRelation1}
\end{equation} Using \eqnref{Iassumption1} and \eqnref{RateRelation1}, we have that
\begin{align}
0> R_i - I(W_{i}) = \frac{R_{1}-I(W_{1})}{1 + \sum_{j=2}^i 2^{l_{j}}} > -\epsilon,
\end{align} for $i=2,\ldots,K$. We first choose $n_{1}=2^{m_{1}}$ for some $m_{1} \in \mathbb{N}$. Then, the choices of $\{R_i\}_{i \in [K]}$ in \eqnref{RateRelation1} guarantees that, for $i=2,\ldots,K$, $n_{i}$ has the form of powers of two since by plugging \eqnref{RateRelation1} into \eqnref{l}, we have:
\begin{align*}
n_{i} &= R_{1}\left(\frac{1}{R_{i}} - \frac{1}{R_{i-1}}\right)n_{1}\\
&= 2^{l_{i}}n_{1} = 2^{m_{1}+l_{i}}.
\end{align*} The rest of proof exactly follows the proof of Theorem~\thmref{capacitytheorem}.

Next we prove the necessity of \eqnref{Iassumption}.  From the definition of the effective rates in \eqnref{EffectiveRates} that any rate-compatible code family must satisfy, we have
 \begin{equation}
 R_i = \frac{R_1}{1 + \sum_{j=2}^i \frac{n_j}{n_1}}   \eqnlabel{RateRelation}
 \end{equation}
 for all $i \in [K]$ by substituting $k = n_1 R_1$.  If the code family is a family rate-compatible polar codes, a polar code can be used in each transmission, and thus we must have that for each $j \in [K]$,
 \begin{equation}
 \frac{n_j}{n_1} = 2^{l_j}   \eqnlabel{LengthLimitation}
 \end{equation}
 for some $l_j \in \mathbb{Z}$ due to the length limitation of polar codes.  Since the set $2^\mathbb{Z} \triangleq \{2^{l}: l \in \mathbb{Z}\}$ is not dense in $\mathbb{R}$, it is not hard to see that those capacities $\{I(W_i)\}_{i\in[K]}$ who can be approached by the rates $\{R_i\}_{i\in[K]}$ that satisfy \eqnref{RateRelation} and \eqnref{LengthLimitation} must satisfy \eqnref{Iassumption}.  To show this rigorously, suppose $\{I(W_i)\}_{i\in[K]}$  cannot be expressed in the form of \eqnref{Iassumption}.  Define
\begin{equation}
\epsilon^* \triangleq \inf_{\{l_j \in \mathbb{Z}\}_{j \in [K]}} \max_{i \in [K]} \left| I(W_i) - \frac{I(W_1)}{1+\sum_{j=2}^{i} 2^{l_j}} \right|.  \eqnlabel{CapacityRelationGap}
\end{equation}
Since $\{I(W_i)\}_{i\in[K]}$ are all distinct and are strictly positive, there must exist a shell $S(\gamma) \triangleq \{x \in \mathbb{R}: \gamma^{-1} < |x| < \gamma\}$ for some sufficiently large $\gamma>0$ such that
\begin{eqnarray}
\epsilon^* 	&=& \inf_{\{l_j \in \mathbb{Z} \cap S(\gamma)\}_{j \in [K]}} \max_{i \in [K]} \left| I(W_i) - \frac{I(W_1)}{1+\sum_{j=2}^{i} 2^{l_j}} \right|   \nonumber \\
		&\stackrel{(a)}{=}& \min_{\{l_j \in \mathbb{Z} \cap S(\gamma)\}_{j \in [K]}} \max_{i \in [K]} \left| I(W_i) - \frac{I(W_1)}{1+\sum_{j=2}^{i} 2^{l_j}} \right|  \nonumber \\
		&\stackrel{(b)}{>}& 0  \nonumber
\end{eqnarray}
where (a) follows since $\mathbb{Z} \cap S(\gamma)$ is a finite (nonempty) set, and (b) follows from the assumption that \eqnref{Iassumption} cannot be satisfied.  Now if there exists a sequence of capacity-achieving rate-compatible polar codes, then there must exist a set of rates $\{R_i\}_{i\in[K]}$ that satisfies \eqnref{RateRelation}, \eqnref{LengthLimitation}, and
\begin{equation}
\max_{i\in[K]} |I(W_i) - R_i| < \frac{\epsilon^*}{2}. \eqnlabel{RateGap}
\end{equation}
It follows that there exists $l_i \in \mathbb{Z}$ for each $i\in[K]$ such that
\begin{eqnarray}
\lefteqn{\max_{i \in [K]} \left| I(W_i) - \frac{I(W_1)}{1+\sum_{j=2}^{i} 2^{l_j}} \right| } \nonumber \\
&\stackrel{(c)}{=}& \max_{i \in [K]} \left| I(W_i) - R_i + \frac{R_1 - I(W_1)}{1+\sum_{j=2}^{i} 2^{l_j}} \right|  \nonumber \\
&\leq& \max_{i \in [K]} \left| I(W_i) - R_i \right|
+ \max_{i \in [K]} \left| R_1 - I(W_1) \right|   \nonumber \\
&\stackrel{(d)}{<}& \frac{\epsilon^*}{2} +  \frac{\epsilon^*}{2} = \epsilon^*
\end{eqnarray}
which contradicts \eqnref{CapacityRelationGap}, where (c) follows from \eqnref{RateRelation} and \eqnref{LengthLimitation}, and (d) follows from \eqnref{RateGap}.

\section{Conclusion} \label{Conclusion}
A method of constructing rate-compatible polar codes that are capacity-achieving with low-complexity sequential decoders is presented.
The proposed code construction  allows for incremental retransmissions at different transmission rates in order to adapt to channel conditions.  The main idea of the construction exploits common characteristics of polar codes optimized for a sequence of degraded channels.
Due to the length limitation of polar codes, the proposed construction cannot support an arbitrary sequence of rates, and we characterize the  rates that can be supported. We then present capacity-achieving  punctured polar codes that provide more flexibility on block length  by controlling a puncturing fraction.  We finally   show that by using such punctured polar codes, the proposed rate-compatible polar code is capacity-achieving for an arbitrary sequence of rates and for any class of degraded channels. The proposed approach uses an optimized polar code to produce the proper amount of incremental redundancy at every HARQ transmission thereby achieving capacity.


\bibliographystyle{IEEEtran}

\end{document}